\documentclass[showpacs,prd,reprint,twocolumn,nofootinbib,showkeys]{revtex4-1}
\pdfoutput=1
\usepackage{graphicx}
\usepackage{dcolumn}
\usepackage{bm}
\usepackage{slashed}
\usepackage{amsmath,graphicx}
\usepackage[colorlinks=true,linktocpage=true,linkcolor=blue,citecolor=blue]{hyperref}
\usepackage{float}
\usepackage{nicefrac}
\usepackage[normalem]{ulem}
\usepackage{amsmath}
\usepackage{subfigure} 
\usepackage{bbold} 
\usepackage[makeroom]{cancel}
\usepackage{stmaryrd} 
\def\CHI{\chi}

\def\be{\begin{equation}}
	\def\ee{\end{equation}}
\newcommand{\bel}[1]{\begin{eqnarray}\label{#1}}
	\newcommand{\eel}{\end{eqnarray}}
\def\barr{\begin{array}}
	\def\earr{\end{array}}
\def\beq{\begin{eqnarray}}
	\def\eeq{\end{eqnarray}}
\def\bfig{\begin{figure}}
	\def\efig{\end{figure}}
\def\lt{\left}
\def\rt{\right}
\newcommand{\nn}{\nonumber}
\newcommand{\f}[2]{\frac{#1}{#2}}
\newcommand{\onehalf}{{\nicefrac{1}{2}}}

\newcommand{\p}{\partial}

\newcommand{\rf}[1]{Eq.~(\ref{#1})}

\newcommand{\rfn}[1]{(\ref{#1})}

\def\a{\alpha}
\def\b{\beta}
\def\g{\gamma}
\def\d{\delta}

\def\LR{\left(} 
\def\RR{\right)}

\def\HP{\hphantom{\alpha}} 

\newcommand{\sh}[1]{\sinh#1}
\newcommand{\ch}[1]{\cosh#1}

\def\half{\frac{1}{2}}

\def\GLW{{\rm GLW}}

\def\nU{n_{(0)}}
\def\eU{\varepsilon_{(0)}}
\def\PU{P_{(0)}}

\newcommand{\lab}[1]{\label{#1}}
\def\nn{\nonumber}

\def\pv{{\boldsymbol p}}
\def\av{{\boldsymbol a}}
\def\bv{{\boldsymbol b}}

\def\Cv{{\boldsymbol C}}

\def\piv{{\boldsymbol \pi}}

\begin{document}
 
    \title{Spin polarization evolution in a boost invariant hydrodynamical background} 
	\author{Wojciech Florkowski}
	\email{wojciech.florkowski@uj.edu.pl}
	 \affiliation{M. Smoluchowski Institute of Physics, Jagiellonian University,  PL-30-348 Krak\'ow, Poland}
	\affiliation{Institute of Nuclear Physics Polish Academy of Sciences, PL-31-342 Krak\'ow, Poland}
	\author{Avdhesh Kumar} 
	\email{avdhesh.kumar@ifj.edu.pl} 
	\affiliation{Institute of Nuclear Physics Polish Academy of Sciences, PL-31-342 Krak\'ow, Poland}
	\author{Radoslaw Ryblewski} 
	\email{radoslaw.ryblewski@ifj.edu.pl}
	\affiliation{Institute of Nuclear Physics Polish Academy of Sciences, PL-31-342 Krak\'ow, Poland}
	\author{Rajeev Singh} 
	\email{rajeev.singh@ifj.edu.pl}
	\affiliation{Institute of Nuclear Physics Polish Academy of Sciences, PL-31-342 Krak\'ow, Poland}
	\date{\today} 
	\bigskip
	\begin{abstract}
	Relativistic hydrodynamic equations for particles with spin $\onehalf$ are used to determine the space-time evolution of the spin polarization in a boost-invariant and transversely homogeneous background. The hydrodynamic approach uses the forms of the energy-momentum and spin tensors based on de Groot, van Leeuwen, and van Weert formalism. Our calculations illustrate how the formalism of hydrodynamics with spin can be used to determine physical observables related to the spin polarization and how the latter can be compared with the experimental data. 
	\end{abstract}
     
\date{\today}

	\pacs{25.75.−q, 24.10.Nz,  24.70.+s}
	
	\keywords{heavy-ion collisions, hydrodynamics, spin polarization, vorticity}
	
\maketitle
%
%

\section{Introduction}
\label{sec:introduction}
The recent measurements of the spin polarization of $\Lambda$-hyperons in relativistic heavy-ion collisions~\cite{STAR:2017ckg, Adam:2018ivw} suggest that the space-time evolution of spin polarization should be included into the hydrodynamic description of such processes. As the hydrodynamic models can be regarded nowadays as a basic tool used for understanding of the space-time evolution of matter created in heavy-ion collisions~\cite{Florkowski:2017olj,Romatschke:2017ejr}, the incorporation of spin dynamics into such models seems to be a natural extension of the standard hydrodynamic approach. Such an extension would offer a new possibility for making comparisons between theory predictions and experimental data.

First steps toward including the spin dynamics in the formalism of relativistic hydrodynamics have been made in Refs.~\cite{Florkowski:2017ruc,Florkowski:2017dyn,Florkowski:2018myy,Becattini:2018duy}, see also the follow-up papers \cite{Florkowski:2017njj,Florkowski:2018amv,Florkowski:2018ual,Florkowski:2018hyw,Florkowski:2019trg} and the review \cite{Florkowski:2018fap}. In this case, the spin dynamics follows solely from the conservation of the angular momentum and other conservation laws, hence, the hydrodynamic equations with spin proposed in Ref.~\cite{Florkowski:2017ruc} can be regarded as a simple extension of the perfect-fluid dynamics.

Other works have dealt so far mainly with the spin polarization of particles at freeze-out~\cite{Becattini:2013fla,Becattini:2013vja,Becattini:2016gvu,Karpenko:2016jyx,Becattini:2017gcx}. In this kind of approach, the basic hydrodynamic quantity giving rise to spin polarization is the thermal vorticity, defined by the expression $\varpi_{\mu \nu} = -\frac{1}{2} (\p_\mu \b_\nu-\p_\nu \beta_\mu)$, where $\beta_\mu$ is the ratio of the fluid flow vector $U_\mu$ and the local temperature $T$, namely $\beta_\mu = U_\mu/T$. A strict relation between the thermal vorticity and the spin polarization tensor $\omega_{\mu\nu}$ (in fact, equality) can be derived for matter in global equilibrium with a rigid rotation \cite{Becattini:2007nd,Becattini:2009wh,Becattini:2015nva}. This reminds the physics situations known from the Einstein -- de Haas and Barnett effects~\cite{dehaas:1915,RevModPhys.7.129}. In the case of heavy-ion collisions we may deal with a similar case in non-central collisions, where a non-vanishing local vorticity perpendicular to the reaction plane is formed~\cite{Liang:2004ph,Liang:2004xn,Betz:2007kg,Becattini:2007sr,Gao:2007bc,Chen:2008wh}. 

In the framework put forward in Ref.~\cite{Florkowski:2017ruc} the spin polarization is described by the spin polarization tensor $\omega_{\mu\nu}$ which is independent of the thermal vorticity. The space-time changes of $\omega_{\mu\nu}$ follow from the conservation laws for angular momentum. Dissipative effects that eventually may bring $\omega_{\mu\nu}$ closer to $\varpi_{\mu \nu}$ are not included. In some sense, the approaches \cite{Florkowski:2017ruc} and \cite{Becattini:2007nd} can be regarded as two extreme cases: in the first case $\omega_{\mu\nu}$ lives its independent live (restricted only by the conservation laws), while in the second case $\omega_{\mu\nu}$ is always constrained to be equal to $\varpi_{\mu \nu}$. One may expect, that in more realistic situations the polarization tensor approaches the thermal vorticity on a characteristic relaxation time-scale~\cite{Becattini:2018duy}. Depending on the magnitude of this relaxation time we may deal with the first or second case. In the future, it would be interesting to explore in more detail the relation of the framework given in Ref.~\cite{Florkowski:2017ruc} to anomalous hydrodynamics \cite{Son:2009tf,Kharzeev:2010gr} and the Lagrangian formulation of hydrodynamics~\cite{Montenegro:2017rbu,Montenegro:2017lvf,Montenegro:2018bcf}.

The hydrodynamic framework worked out in Refs.~\cite{Florkowski:2017ruc,Florkowski:2017dyn} is based on the specific forms of the energy-momentum and spin tensors. These forms have been chosen in such a way as to obtain the simplest possible description which is self-consistent from the thermodynamic and hydrodynamic point of view. A more recent work has demonstrated, however, that other forms of the energy-momentum and/or spin tensors should be used if we want to connect them with the underlying kinetic theory~\cite{Florkowski:2018ahw}, see also   \cite{Kumar:2018iud,Kumar:2018lok}. As a matter of fact, these forms agree with those introduced by de Groot, van Leeuwen,  and van Weert (GLW) in Ref.~\cite{DeGroot:1980dk}. It has been also shown in~Ref.~\cite{Florkowski:2018ahw} that the GLW forms are connected with the canonical expressions (given through the Noether theorem) via the so-called pseudo-gauge transformation~\cite{Hehl:1976vr,Becattini:2018duy,Florkowski:2018fap}. In view of this fact, we have decided to consider here the case where the hydrodynamics with spin is formulated with the GLW forms of the energy-momentum and spin tensors. 

Another important limitation of the formulation~\cite{Florkowski:2017ruc,Florkowski:2017dyn} is that it does not allow for arbitrary large values of the polarization tensor~\cite{Florkowski:2018fap}. Therefore, in the present approach we restrict ourselves to the leading-order expressions in the polarization tensor  $\omega_{\mu\nu}$.~\footnote{The spin polarization tensor $\omega_{\mu\nu}$ is dimensionless, it can be defined as the ratio of the spin chemical potential $\Omega_{\mu\nu}$ and the temperature $T$, $\omega_{\mu\nu} = \Omega_{\mu\nu}/T$.} We note that, a priori, we cannot say if the hydrodynamic equations do not lead to instabilities which make higher-order terms in $\omega_{\mu\nu}$ important. This should be individually checked for each form of the initial conditions. 

The conclusion from the points discussed above is that the most convincing framework for hydrodynamics with spin is that based on the GLW forms of the energy-momentum and spin tensors, combined with the linear expansion in $\omega_{\mu\nu}$. Since at the moment no solutions of such a scheme are known, the purpose of this paper is to explore the simplest, boost-invariant expansion geometry known as the Bjorken expansion \cite{Bjorken:1982qr},  and to look for the consequences of the hydrodynamic scheme introduced in this way. In addition, we assume that the systems studied here are transversely homogeneous.  

An attractive feature of our scheme is that the terms linear in $\omega_{\mu\nu}$ appear only in the spin tensor. Hence, the conservation of energy and linear momentum can be analyzed in exactly the same way as in standard hydrodynamics and, subsequently, the spin evolution in a given hydrodynamic background can be determined. 

The study presented in this work can be used as a practical illustration as well as a check of the theoretical scheme defined above. The latter consists of four distinct steps: i) solving the standard perfect-fluid hydrodynamic equations without spin, ii) determination of the spin evolution in the hydrodynamic background, iii) determination of the Pauli-Luba\'nski (PL) vector on the freeze-out hypersurface, and, finally, iv) calculation of the spin polarization of particles in their rest frame. The spin polarization obtained in this way is a function of the three-momenta of particles and can be directly compared with the experiment. 
 
In the context of the recent experiments, probably the most interesting issue is the determination of the longitudinal (i.e., along the beam axis) polarization of $\Lambda$ and $\bar{\Lambda}$. This observable was first discussed by Jacob and Rafelski \cite{Jacob:1987sj,Jacob:1988dc}, however, the first heavy-ion collision experiments in Dubna \cite{Anikina:1984cu}, at CERN \cite{Bartke:1990cn} and BNL \cite{Abelev:2007zk} reported negative results. A breakthrough came when the $\Lambda(\bar{\Lambda})$-–hyperon spin polarization was measured very recently by the STAR collaboration~\cite{STAR:2017ckg, Adam:2018ivw}.

Interestingly, the STAR measurement also shows a quadrupole dependence of the longitudinal polarization with respect to the reaction plane \cite{Niida:2018hfw}. It turns out, that this behavior cannot be reproduced by the current model calculations  \cite{Becattini:2017gcx} which assume that spin polarization tensor is equal to the thermal vorticity, although the difference resides mainly in the sign of the polarization. If this difference persists, it may suggest that, indeed, the spin polarization evolves independently from the thermal vorticity. Interestingly, very recent simulations~\cite{Sun:2018bjl} based on the chiral kinetic theory have been able to explain the longitudinal polarization in the scenario where $\omega_{\mu\nu} \neq \varpi_{\mu\nu}$. 

Due to the simplified geometry, the hydrodynamical model described herein cannot describe properly the longitudinal polarization. Nevertheless, our calculations demonstrate how the formalism of hydrodynamics with spin can be used to determine spin observables and how they can be compared with the experimental data. In this way, the calculations presented herein set the stage for more realistic calculations. 

\bigskip
{\it Notation and conventions:} 
 The metric tensor is taken as  $g_{\mu\nu} =  \hbox{diag}(+1,-1,-1,-1)$. The scalar product of two four-vectors $a^{\mu}$ and $b^{\mu}$ reads $a \cdot b =a^{\mu}b_{\mu}= g_{\mu \nu} a^\mu b^\nu = a^0 b^0 - \av \cdot \bv$, where bold font is used to denote three-vectors. For the  Levi-Civita tensor $\epsilon^{\mu\nu\rho\sigma}$ the convention $\epsilon^{0123} = -\epsilon_{0123} =+1$ is used. The Lorentz invariant measure in the momentum space is represented by $dP = \frac{d^3p}{(2 \pi )^3 E_p}$, where $E_p = \sqrt{m^2 + \pv^2}$ and $p^\mu = (E_p, \pv)$ are the on-mass-shell particle energy and the particle four-momentum, respectively. The square brackets are used to denote antisymmetrization with respect to a pair of indices, say $\mu$ and $\nu$, for example, $A^{[\mu \nu]} =   \left(A^{\mu\nu} - A^{\nu\mu} \right)/2$.  Any dual tensor, obtained by contracting a rank-two antisymmetric tensor with the  Levi-Civita tensor and dividing by a factor of two is represented by a symbol tilde over it. For example, the dual tensor to ${t}^{\mu\nu}$ is defined as
\bel{eq:dual}
\tilde{t}^{\mu\nu} = \f{1}{2} \epsilon^{\mu\nu\alpha\beta}  {t}_{\alpha\beta}.
\eel
The inverse transformation is
\bel{eq:dualdual}
{t}^{\mu\nu} =-\f{1}{2} \epsilon^{\mu\nu\alpha\beta}  \tilde{t}_{\alpha\beta}.
\eel
Throughout the text, natural units {\it i.e.} $c = \hbar = k_B~=1$ are used.
%
\section{Spin polarization tensor and Perfect fluid hydrodynamics for particles with spin 1/2}
\subsection{Spin polarization tensor}
The spin polarization tensor $\omega_{\mu\nu}$ is antisymmetric and can be defined by the four-vectors $\kappa^\mu$ and $\omega^\mu$ \cite{Florkowski:2017ruc}, 
\beq
\omega_{\mu\nu} &=& \kappa_\mu U_\nu - \kappa_\nu U_\mu + \epsilon_{\mu\nu\a\b} U^\a \omega^{\b}, \lab{spinpol1}
\eeq
where $U^\mu$ is the flow four-vector. It is important to note that any part of the four-vectors $\kappa^{\mu}$ and $\omega^{\mu}$ which is parallel to  $U^{\mu}$ does not contribute to the right-hand side of~Eq.~(\ref{spinpol1}). Hence, we can assume that $\kappa^{\mu}$ and $\omega^{\mu}$  satisfy the orthogonality conditions~\footnote{Six independent components of $\kappa^{\mu}$ and $\omega^{\mu}$ define six independent components of the antisymmetric tensor $\omega_{\mu\nu}$.}
\beq
\kappa\cdot U = 0, \quad \omega \cdot U = 0  \lab{ko_ortho}.
\eeq
Using these constraints, we can express $\kappa_\mu$ and $\omega_\mu$ in terms of $\omega_{\mu\nu}$, namely
\beq
\kappa_\mu= \omega_{\mu\a} U^\a, \quad \omega_\mu = \half \epsilon_{\mu\a\b\g} \omega^{\a\b} U^\g. \lab{eq:kappaomega}
\eeq
%
\subsection{Perfect fluid hydrodynamics for particles with spin 1/2}
In this section we define the hydrodynamic equations for particles with spin $\onehalf$. Having in mind our earlier remarks about the GLW formalism and expansion in $\omega_{\mu\nu}$, we ignore spin degrees of freedom in the conservation laws for charge, energy and linear momentum. Consequently, the polarization tensor is included only in the conservation of angular momentum.
%
\subsubsection{Conservation of charge}
The conservation law of charge current~\footnote{The charge may represent here any of the conserved charges such as the electric charge or baryon number. } is expressed by the standard expression
\bel{eq:Ncon}
  \p_\alpha N^\alpha(x)  = 0,
\eel
where
\bel{eq:Nmu}
N^\alpha = n U^\alpha
\eel
and \cite{Florkowski:2017ruc}
\bel{nden}
n = 4 \, \sinh(\xi)\, \nU(T).
\eel
Here we assume the equation of state of an ideal relativistic gas of classical massive particles (and antiparticles) with spin $\onehalf$. The quantity $\nU(T)$ defines the number density of spinless and neutral massive Boltzmann particles, 
\bel{n0}
\nU(T) = \langle p\cdot U\rangle_0 \, ,
\eel 
with $\langle\, \cdots \rangle_0$ denoting a thermal average  
\bel{thermal_av}
\langle\, \cdots \rangle_0 \equiv \int{dP}  \,(\cdots) \,  e^{- \beta \cdot p}.
\eel
The factor $4 \, \sinh(\xi) = 2 \left(e^\xi - e^{-\xi} \right)$ in~\rf{nden} accounts for spin degeneracy and presence of both particles and antiparticles in the system. The variable $\xi$ is the ratio of the chemical potential $\mu$ and the temperature, $\xi=\mu/T$.

Using \rf{thermal_av} in \rf{n0} and carrying out the momentum integrals one obtains the well-known result (see, e.g., Ref.~\cite{Florkowski:2010zz})
\beq
\nU(T) &=&  \f{1}{2\pi^2}\, T^3 \, \hat{m}^2 K_2\left( \hat{m}\right), \label{polden}
\eeq
where $T$ is the temperature, $\hat{m}\equiv m/T$ is the ratio of the particle mass and the temperature, and  $K_2\left( \hat{m}\right)$ denotes the modified Bessel function. 
%
\subsubsection{Conservation of energy and linear momentum}
The conservation of energy and linear momentum is expressed by the equation
\bel{eq:Tcon}
\p_\a T^{\a\b}_{\rm GLW}(x) = 0,
\eel
where the energy-momentum tensor $T^{\a\b}_{\rm GLW}$ has the perfect-fluid form 
\bel{Tmn}
T^{\a\b}_{\rm GLW} &=& (\varepsilon + P ) U^\a U^\b - P g^{\a\b}
\eel
with the energy density and pressure given by
\bel{enden}
\varepsilon = 4 \, \cosh(\xi) \, \eU(T)
\eel
and 
\bel{prs}
P = 4 \, \cosh(\xi) \, \PU(T),
\eel
respectively. In analogy to the density $\nU(T)$, the auxiliary quantities $\eU(T)$ and $\PU(T)$ are defined as $\eU(T) = \langle(p\cdot U)^2\rangle_0$ and $\PU(T) = -(1/3) \langle  p\cdot p - (p\cdot U)^2   \rangle_0$. For an ideal relativistic gas of classical massive particles one finds~\cite{Florkowski:2010zz}
\beq
\eU(T) &=& \f{1}{2\pi^2} \, T^4 \, \hat{m} ^2
 \Big[ 3 K_{2}\left( \hat{m} \right) + \hat{m}  K_{1} \left( \hat{m}  \right) \Big],  \label{eneden}\\
\PU(T) &=& T \, \nU(T)  . \label{P0}
\eeq

At this point it is important to notice that Eqs.~(\ref{eq:Ncon}) and (\ref{eq:Tcon}) form a closed system of five equations for five unknown functions: $\xi$, $T$, and three independent components of $U^\mu$. They are nothing else but the perfect-fluid equations which should be solved in the first step in order to define a hydrodynamic background for the spin dynamics.
%
\subsubsection{Conservation of angular momentum}
Since the energy-momentum tensor used in the GLW framework is symmetric, the conservation of the angular momentum implies the conservation of its spin part, i.e., of the spin tensor. Thus, in the GLW formalism we use the formula~\cite{Florkowski:2018ahw}
\beq
\p_\a S^{\a , \beta \gamma }_{\rm GLW}(x)&=& 0,
\label{eq:SGLWcon}
\eeq
where the GLW spin tensor in the leading order of  $\omega_{\mu\nu}$ is given by the expression \cite{Florkowski:2018ahw}
\beq
S^{\alpha , \beta \gamma }_{\rm GLW}
&=&  {\cal C} \left( n_{(0)}(T) U^\alpha \omega^{\beta\gamma}  +  S^{\a, \b\g}_{\Delta\GLW} \right),
\label{eq:SGLW}
\eeq
with ${\cal C} = \ch(\xi)$.
Here, the auxiliary tensor $S^{\a, \b\g}_{\Delta\GLW}$ is defined as~\cite{Florkowski:2017dyn}
\beq
S^{\a, \b\g}_{\Delta\GLW} 
&=&  {\cal A}_{(0)} \, U^\a U^\d U^{[\b} \omega^{\g]}_{\HP\d} \lab{SDeltaGLW} \\
&& \hspace{-0.5cm} + \, {\cal B}_{(0)} \, \Big( 
U^{[\b} \Delta^{\a\d} \omega^{\g]}_{\HP\d}
+ U^\a \Delta^{\d[\b} \omega^{\g]}_{\HP\d}
+ U^\d \Delta^{\a[\b} \omega^{\g]}_{\HP\d}\Big),
\nn
\eeq
where
\beq 
{\cal B}_{(0)} &=&-\frac{2}{\hat{m}^2}  \frac{\varepsilon_{(0)}(T)+P_{(0)}(T)}{T}=-\frac{2}{\hat{m}^2} s_{(0)}(T)\label{coefB}
\eeq
and
\beq
{\cal A}_{(0)} &=&\frac{6}{\hat{m}^2} s_{(0)} +2 n_{(0)} (T) = -3{\cal B}_{(0)} +2 n_{(0)}(T).
\label{coefA}
\eeq
In the following we shall use yet another decomposition of the spin tensor~\rfn{eq:SGLW}, namely
\beq
S^{\a, \b\g}_\GLW 
&=& {\cal A}_1 U^\a \omega^{\b\g} 
+ {\cal A}_2 U^\a U^{[\b} \kappa^{\g]} \nn \\
&&
+  {\cal A}_3 (U^{[\b} \omega^{\g]\a} + g^{\a[\b} \kappa^{\g]}) ,
\lab{SGLW2}
\eeq
where
\beq
{\cal A}_1 &=& {\cal C} \LR \nU -  {\cal B}_{(0)} \RR \label{A1} ,\\ 
{\cal A}_2 &=& {\cal C} \LR {\cal A}_{(0)} - 3{\cal B}_{(0)} \RR  \label{A2} , \\ 
{\cal A}_3 &=&  {\cal C}\, {\cal B}_{(0)}\label{A3}. 
\eeq
%
%
\section{Boost-invariant flow and spin polarization tensor}
\subsection{Implementation of boost invariance}
For systems which are boost invariant and transversely homogeneous, it is useful to introduce a local basis consisting of the following four-vectors:
\beq
U^\a &=& \frac{1}{\tau}\LR t,0,0,z \RR = \LR \ch(\eta), 0,0, \sh(\eta) \RR, \nn \\
X^\a &=& \LR 0, 1,0, 0 \RR,\nn\\
Y^\a &=& \LR 0, 0,1, 0 \RR, \nn\\
Z^\a &=& \frac{1}{\tau}\LR z,0,0,t \RR = \LR \sh(\eta), 0,0, \ch(\eta) \RR. 
\lab{BIbasis}
\eeq
Here $\tau = \sqrt{t^2-z^2}$ is the longitudinal proper time, while $\eta =  \ln((t+z)/(t-z))/2$ is the space-time rapidity. The four-vectors (\ref{BIbasis}) are boost invariant, which means that after performing a Lorentz boost $L^\mu_{\,\,\,\nu}$ along the $z$-axis, their new components $V^{\prime \mu}$ at the new space-time points $x^\prime$ agree with the original components $V^\mu$ at $x^\prime$, $V^{\prime \mu}(x^\prime) = L^\mu_{\,\,\,\nu} V^\nu(x) =  V^\mu(x^\prime)$~\cite{Florkowski:2010zz}. For scalar functions of space and time coordinates, such as $T(x)$ or $\xi(x)$, the boost invariance implies that they may depend only on the variable~$\tau$, hence, $T=T(\tau)$ and $\xi=\xi(\tau)$.

The four-vector $U^\a$ is time-like and normalized to unity, while the four-vectors $X^\a, Y^\a$ and $Z^\a$ are space-like and orthogonal to $U^\a$ as well as to each other,
\begin{eqnarray}
 && \,\,U \cdot U = 1\label{UU}\\
X \cdot X \,\,&=& \,\, Y \cdot Y \,\,=\,\, Z \cdot Z \,\,=\,\, -1, \\ \label{XXYYZZ}
X \cdot U\,\, &=& \,\,Y \cdot U \,\,=\,\, Z \cdot U \,\,=\,\, 0,  \\ \label{XYZU}
X \cdot Y\,\, &=& \,\, Y \cdot Z \,\,=\,\, Z \cdot X \,\,=\,\, 0.  \label{XYYZZX}
\end{eqnarray}
As we have mentioned above, we identify $U^\a$ with the flow vector of matter. The local rest frame (of the fluid element) is defined as the frame where $U^\a = (1,0,0,0)$.

In the following, we shall use also derivatives with respect to $\tau$ and $\eta$. They are connected with the standard derivatives through the expression
{\[
\begin{bmatrix}
\p_t \\
\p_x \\
\p_y \\
\p_z \end{bmatrix}=
\begin{bmatrix}
\ch(\eta) & 0 & 0 & -\sh(\eta)  \\
0 & 1 & 0 & 0 \\
0 & 0 & 1 & 0  \\
-\sh(\eta) & 0 & 0 & \ch(\eta) \\
\end{bmatrix}\begin{bmatrix}
\p_\tau \\
\p_x \\
\p_y \\
\frac{1}{\tau}\p_\eta \end{bmatrix}.
\]}
Using this transformation one can find useful relations:
\beq
 \p \cdot U = \frac{1}{\tau}, \quad&&\quad U \cdot \p = \p_\tau \equiv \dot{(\HP)}, \\
\p \cdot X = 0, \quad&&\quad X \cdot \p = \p_x,\\
\p \cdot Y = 0, \quad&&\quad Y \cdot \p = \p_y,\\
\p \cdot Z = 0, \quad&&\quad Z \cdot \p = \frac{1}{\tau} \partial_{\eta}.
\eeq

Using the basis \rfn{BIbasis}, one can introduce the following representation of the vectors $\kappa^{\mu}$ and $\omega^{\mu}$ defined by~\rf{eq:kappaomega}
\beq
\kappa^\a &=&  C_{\kappa X} X^\a + C_{\kappa Y} Y^\a + C_{\kappa Z} Z^\a, \lab{eq:k_decom}\\
\omega^\a &=&  C_{\omega X} X^\a + C_{\omega Y} Y^\a + C_{\omega Z} Z^\a. \lab{eq:o_decom}
\eeq
Here, the scalar coefficients ${C}_{\kappa X}$, 
${C}_{\kappa Y}$, ${C}_{\kappa Z}$, ${C}_{\omega X}$, ${C}_{\omega Y}$, and ${C}_{\omega Z}$ (below we generically refer to them as to the $C$ coefficients) are functions of the proper time $\tau$ only.  It is important to note that due to the orthogonality conditions~\rfn{ko_ortho}, there are no terms proportional to $U^\a$ in Eqs.~\rfn{eq:k_decom} and \rfn{eq:o_decom}. 

Substituting Eqs.~\rfn{eq:k_decom} and \rfn{eq:o_decom} into \rf{spinpol1} we obtain a boost-invariant expression for the spin polarization tensor $\omega_{\mu\nu}$, 
\beq
\omega_{\mu\nu} &=& C_{\kappa Z} (Z_\mu U_\nu - Z_\nu U_\mu) \label{eq:omegamunu} \\
&& + C_{\kappa X} (X_\mu U_\nu - X_\nu U_\mu) \nonumber \\
&& + C_{\kappa Y} (Y_\mu U_\nu - Y_\nu U_\mu) \nonumber \\
&& + \, \epsilon_{\mu\nu\alpha\beta} U^\alpha (C_{\omega Z} Z^\beta + C_{\omega X} X^\beta + C_{\omega Y} Y^\beta). \nn
\eeq
In the plane $z=0$ we find
\beq
\omega_{\mu\nu} =
\begin{bmatrix}
0 & C_{\kappa X} & C_{\kappa Y} & C_{\kappa Z} \\
-C_{\kappa X} & 0 & -C_{\omega Z} & C_{\omega Y} \\
-C_{\kappa Y} & C_{\omega Z} & 0 & -C_{\omega X}  \\
-C_{\kappa Z} & -C_{\omega Y} & C_{\omega X} & 0 \\
\end{bmatrix}.
\eeq
Finally, using \rf{eq:omegamunu} we obtain a boost-invariant expression for the spin tensor $S^{\a , \b\g}_{\GLW}$.
%
%
\subsection{Spin and orbital angular momentum of a boost-invariant fire-cylinder}
\begin{figure}[t]
\centering
\includegraphics[width=0.45\textwidth]{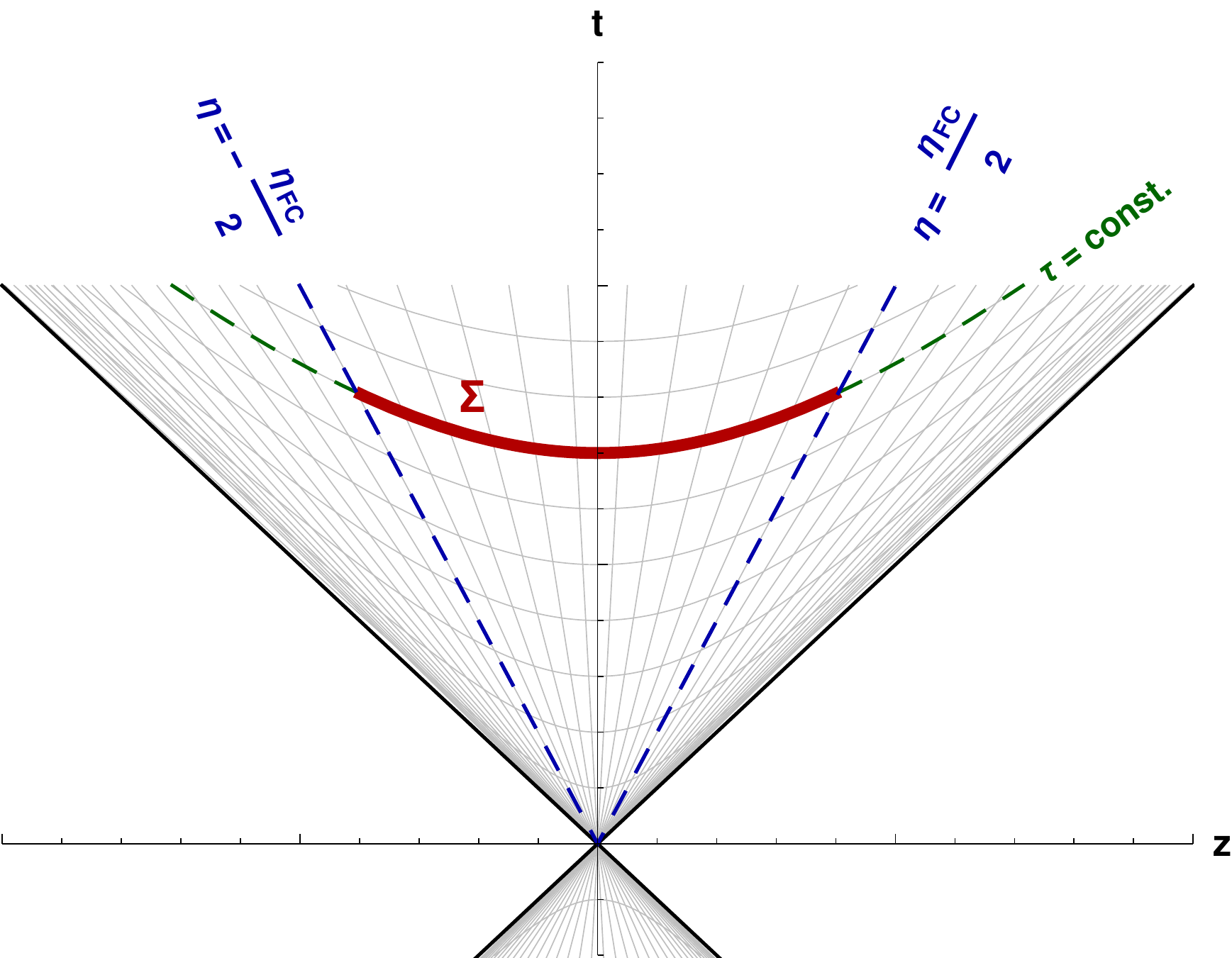}
\caption{(Color online) The hypersurface of the boost-invariant fire-cylinder.}
\label{fig:fc}
\end{figure}

In order to get more insight into the physics interpretation of the coefficients $C$, we consider now a boost-invariant fire-cylinder (FC) occupying the space-time region defined by the conditions: $\tau=$~const, $-\eta_{\rm FC}/2 \leq \eta \leq +\eta_{\rm FC}/2$, and $\sqrt{x^2+y^2} \leq R$, see Fig.~\ref{fig:fc}. In this case, a small space-time element of the fire-cylinder, $\Delta \Sigma _{\lambda }$, can be defined by the formula
\beq
\Delta \Sigma _{\lambda } &=& U_{\lambda }\, dx dy\, \tau  d\eta. \lab{sig} 
\eeq

The spin part of the total angular momentum contained in the fire-cylinder is
\beq
S^{\mu\nu}_{\rm FC} &=& \int
\Delta \Sigma _{\lambda } S^{\lambda,\mu\nu}_{\rm GLW} =
\int
dx dy \,\tau  d\eta \, U_{\lambda } S^{\lambda,\mu\nu}_{\rm GLW} \nn \\ &=& \pi R^2 \tau \int\limits_{-\eta_{\rm FC}/2}^{+\eta_{\rm FC}/2} d\eta \, U_{\lambda } S^{\lambda,\mu\nu}_{\rm GLW} .
\label{SAM}
\eeq
Using Eqs.~\rfn{SGLW2},  \rfn{eq:k_decom},  \rfn{eq:o_decom} and \rfn{sig} in \rf{SAM}, one can obtain the following expression for $S^{\mu\nu}_{\rm FC}$, 
\beq
S^{\mu\nu}_{\rm FC}&=& \pi R^2 \tau \int\limits_{-\eta_{FC}/2}^{+\eta_{\rm FC}/2} d\eta   \left[ \vphantom{\int}  \right. {\cal A}_\kappa \big[ C_{\kappa X} \left(U^{\nu } X^{\mu }-U^{\mu } X^{\nu }\right)\nn\\&&+C_{\kappa Y} \left(U^{\nu } Y^{\mu }-U^{\mu } Y^{\nu }\right)+C_{\kappa Z} \left(U^{\nu } Z^{\mu }-U^{\mu } Z^{\nu }\right)\big] \nn\\&&+{\cal A}_1 \epsilon ^{\mu \nu \delta \chi } U_{\delta } \left( C_{\omega X} X_{\chi }+C_{\omega Y} Y_{\chi }+C_{\omega Z} Z_{\chi } \right) \left. \vphantom{\int}  \right], \nn \\
\label{SAMBI}
\eeq
where ${\cal A}_\kappa \equiv {\cal A}_1+ ({\cal A}_2+2{\cal A}_3)/2$. In the next step,  using \rf{BIbasis}  one can evaluate all the components of $S^{\mu\nu}_{\rm FC}$, which are given by the following antisymmetric matrix
\begin{equation}
S^{\mu\nu}_{\rm FC} = -S^{\nu\mu}_{\rm FC} =
\begin{bmatrix}
\vspace{0.2cm}
0 & S^{01}_{\rm FC}  & S^{02}_{\rm FC} & S^{03}_{\rm FC} \\ \vspace{0.2cm}
-S^{01}_{\rm FC} & 0 & S^{12}_{\rm FC} & S^{13}_{\rm FC} \\ \vspace{0.2cm}
-S^{02}_{\rm FC} & -S^{12}_{\rm FC} & 0 & S^{23}_{\rm FC} \\ \vspace{0.1cm}
-S^{03}_{\rm FC} & -S^{13}_{\rm FC}  & -S^{23}_{\rm FC} & 0 \\
\end{bmatrix}, \lab{LT}
\end{equation}
where (with contravariant indices replaced by the covariant ones):
\beq
S_{01}^{\rm FC}&=&2\pi R^2 \tau  \,{\cal A}_\kappa C_{\kappa X} \sinh(\eta_{\rm FC}/2), \nn\\
S_{02}^{\rm FC}&=&2\pi R^2 \tau  \,{\cal A}_\kappa C_{\kappa Y} \sinh(\eta_{\rm FC}/2), \nn\\
S_{03}^{\rm FC}&=&\pi R^2 \tau \, {\cal A}_\kappa C_{\kappa Z} \, \eta_{\rm FC}, \nn\\
S_{23}^{\rm FC}&=&-2 \pi R^2 \tau \,{\cal A}_1 C_{\omega X} \sinh(\eta_{\rm FC}/2),  \nn\\
S_{13}^{\rm FC}&=&2\pi R^2 \tau \,{\cal A}_1  C_{\omega Y} \sinh(\eta_{\rm FC}/2), \nn \\
S_{12}^{\rm FC}&=&-\pi R^2 \tau   \,{\cal A}_1 C_{\omega Z} \, \eta_{\rm FC}.
\eeq
We thus see that the coefficients $C$ directly define  different components of the total spin angular momentum of the boost-invariant fire-cylinder.

At this place it is also interesting to discuss the orbital contribution to the total angular momentum of the fire-cylinder. It is given by the expression
\beq
L^{\mu\nu}_{\rm FC} =\int
\Delta \Sigma _{\lambda } L^{\lambda,\mu\nu} =\int
\Delta \Sigma _{\lambda } \lt(x^{\mu}T^{\lambda\nu}_{\rm GLW}
-x^{\nu}T^{\lambda\mu}_{\rm GLW} \rt).
\nn\\\label{OAM}
\eeq
Using Eqs.~\rfn{Tmn} and \rfn{sig} in \rf{OAM} we can write,
\beq
L^{\mu\nu}_{\rm FC} =\int
dx dy\, \tau  d\eta \,\varepsilon \lt(x^{\mu}U^{\nu}_{\rm GLW}-x^{\nu}U^{\mu}_{\rm GLW}\rt).
\eeq
Substituting $U^{\mu}$ from \rf{BIbasis} into this equation, one can easily show that for our system
\beq
L^{\mu\nu}_{\rm FC} =0.
\eeq
Thus, the only finite contribution to the total angular momentum comes from the spin part. 
%
\subsection{Boost-invariant forms of the conservation laws}
Using \rf{eq:Nmu} in \rf{eq:Ncon}, the conservation law for charge can be written as
\beq
U^{\a}\p_{\a}n+n\p_{\a}U^{\a}=0.
\eeq
Thus, for the Bjorken flow defined above we obtain
\beq
\dot{n}+\frac{n}{\tau}=0.\lab{eq:charge}
\eeq
This equation has a simple scaling solution $n = n_0 \tau_0/\tau$, where $n_0$ is the initial density ($n$ at $\tau=\tau_0$).

Contracting \rf{eq:Tcon} with $U_{\b}$ and $\Delta^{\mu}_{\b} = g^{\mu}_{\b} - U^\mu U_\beta$, respectively, and then using \rf{Tmn}, we obtain two equations
\beq
U^{\a}\p_{\a}\varepsilon+(\varepsilon+P)\p_{\a}U^{\a}&=&0,\lab{eq:encons}\\
(\varepsilon+P)U^{\a}\p_{\a}U^{\mu}-\Delta^{\mu\a}\p_{\alpha}P&=&0.\lab{eq:momcons}
\eeq
Equation \rfn{eq:encons} is equivalent to the entropy conservation, while \rf{eq:momcons} is a relativistic generalization of the Euler equation. For the Bjorken flow geometry, the latter can be written simply as
\beq
\dot{\varepsilon}+\frac{(\varepsilon+P)}{\tau}=0.\lab{eq:en}
\eeq

Using Eqs.~\rfn{eq:k_decom} and \rfn{eq:o_decom} in \rf{SGLW2} and subsequently in \rf{eq:SGLWcon}, and contracting the resulting tensor equation with $U_\b X_\g$, $U_\b Y_\g$, $U_\b Z_\g$,  $Y_\b Z_\g$, $X_\b Z_\g$ and $X_\b Y_\g$, respectively, the following set of the evolution equations for the coefficients $C$ can be obtained:
\begin{widetext}
\begin{equation}
\begin{bmatrix}
\cal{L}(\tau) & 0 & 0 & 0 & 0 & 0 \\
0 & \cal{L}(\tau) & 0 & 0 & 0 & 0 \\
0 & 0 & \cal{L}(\tau) & 0 & 0 & 0 \\
0 & 0 & 0 & \cal{P}(\tau) & 0 & 0 \\
0 & 0 & 0 & 0 & \cal{P}(\tau)  & 0 \\
0 & 0 & 0 & 0 & 0 & \cal{P}(\tau)\end{bmatrix}
\begin{bmatrix}
\Dot{C}_{\kappa X} \\
\Dot{C}_{\kappa Y} \\
\Dot{C}_{\kappa Z} \\
\Dot{C}_{\omega X} \\
\Dot{C}_{\omega Y} \\
\Dot{C}_{\omega Z} \end{bmatrix}=\begin{bmatrix}
{\cal{Q}}_1(\tau) & 0 & 0 & 0 & 0 & 0 \\
0 & {\cal{Q}}_1(\tau) & 0 & 0 & 0 & 0 \\
0 & 0 & {\cal{Q}}_2(\tau) & 0 & 0 & 0 \\
0 & 0 & 0 & {\cal{R}}_1(\tau) & 0 & 0 \\
0 & 0 & 0 & 0 & {\cal{R}}_1(\tau)  & 0 \\
0 & 0 & 0 & 0 & 0 & {\cal{R}}_2(\tau)\end{bmatrix}
\begin{bmatrix}
{C}_{\kappa X} \\
{C}_{\kappa Y} \\
{C}_{\kappa Z} \\
{C}_{\omega X} \\
{C}_{\omega Y} \\
{C}_{\omega Z} \end{bmatrix}, \label{cs}
\end{equation}
\end{widetext}
where
\beq
{\cal L}(\tau)&=&{\cal A}_1-\frac{1}{2}{\cal A}_2-{\cal A}_3,\nn\\
{\cal P}(\tau)&=&{\cal A}_1,\nn\\
{\cal{Q}}_1(\tau)&=&-\left[\dot{{\cal L}}+\frac{1}{\tau}\left( {\cal L}+ \frac{1}{2}{\cal A}_3\right)\right],\nn\\
 {\cal{Q}}_2(\tau)&=&-\left(\dot{{\cal L}}+\frac{{\cal L}}{\tau}   \right),\nn\\
  {\cal{R}}_1(\tau)&=&-\left[\Dot{\cal P}+\frac{1}{\tau}\left({\cal P} -\frac{1}{2} {\cal A}_3 \right) \right],\nn\\
 {\cal{R}}_2(\tau)&=&-\left(\Dot{{\cal P}} +\frac{{\cal P}}{\tau}\right).
 \label{LPQR}
 \eeq
 
 Interestingly, we find that all the coefficients $C$ evolve independently. We also find that the coefficients ${C}_{\kappa X}$ and ${C}_{\kappa Y}$ (and similarly ${C}_{\omega X}$ and ${C}_{\omega Y}$) obey the same differential equations. This is caused by the rotational invariance in the transverse plane. Moreover, since Eqs.~\rfn{cs} are uniform, each of the coefficients $C$ remains equal to zero if its initial value is zero.
 %
\subsection{Numerical results}

In this section we present numerical solutions of Eqs.~\rfn{eq:charge}, \rfn{eq:en} and \rfn{cs}. As stated above, we first solve Eqs.~\rfn{eq:charge} and \rfn{eq:en}. In this way we determine proper-time dependence of the temperature $T$ and chemical potential $\mu$ (note that $\xi = \mu/T$). If the functions $T(\tau)$ and $\mu(\tau)$ are known, one can easily determine the functions ${\cal L}$, ${\cal P}$, ${\cal Q}$ and ${\cal R}$ appearing on the left- and right-hand sides of \rf{cs}. Then, it is also possible to find the time dependence of the coefficients $C$ which define the polarization tensor. 

In order to address physical situations similar to those studied experimentally, we consider a baryon rich matter with the initial baryon chemical potential $\mu_0=800$~MeV and the initial temperature $T_0=155$~MeV. The particle mass is taken to be equal to that of the $\Lambda$-hyperon, $m~=~1116$~MeV. The initial proper time is $\tau_0=1$~fm, and we continue the hydrodynamic evolution till the final time $\tau_f=$~10~fm.

\begin{figure}[ht!]
\centering
\includegraphics[width=0.45\textwidth]{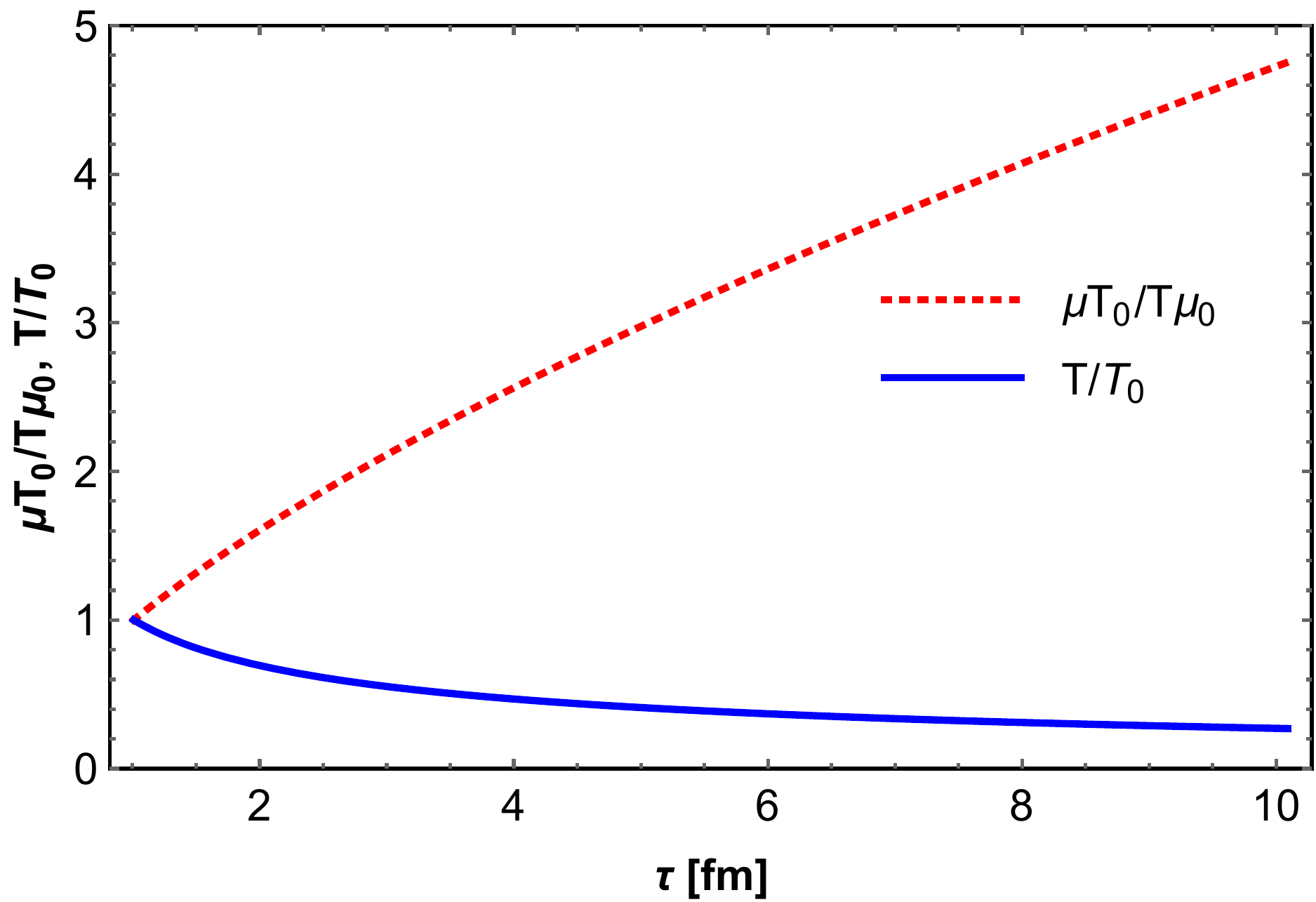}
\caption{Proper-time dependence of $T$ divided by its initial value $T_0$ (solid line) and the ratio of baryon chemical potential $\mu$ and temperature $T$  rescaled by the initial ratio $\mu_0/T_0$ (dotted line) for a boost-invariant one-dimensional expansion.}
\label{fig:Tmu}
\end{figure}

In Fig.~\ref{fig:Tmu} we show the proper-time dependence of the temperature and baryon chemical potential obtained from Eqs.~\rfn{eq:charge} and \rfn{eq:en}. We reproduce well established results that the temperature decreases with $\tau$, while the ratio of the chemical potential and temperature grows up. We note that in the case of massless particles the Bjorken scenario predicts a constant $\mu/T$ ratio and $T = T_0 (\tau_0/\tau)^{1/3}$. 
\begin{figure}[ht!]
\centering
\includegraphics[width=0.45\textwidth]{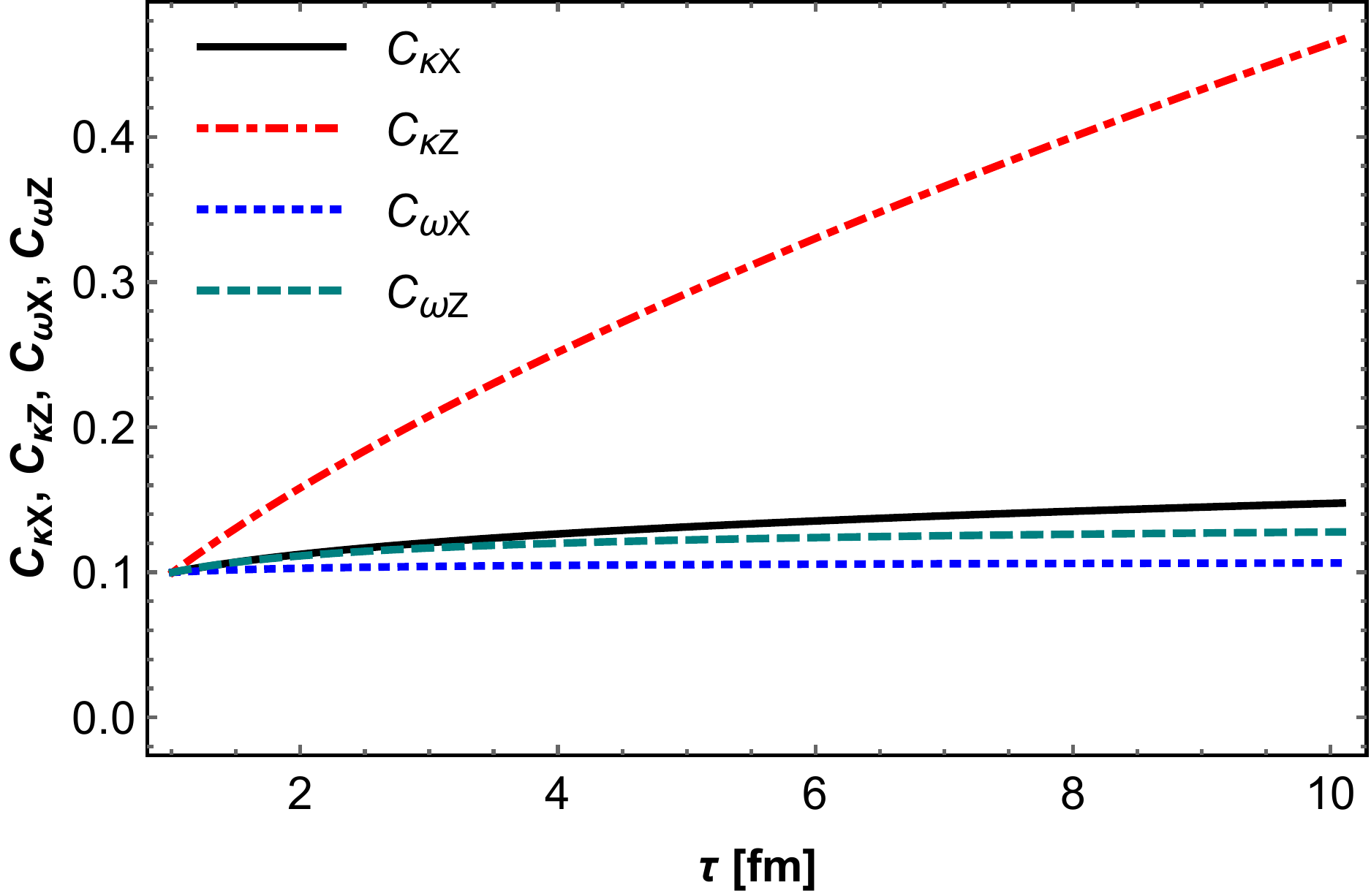}
\caption{Proper-time dependence of the coefficients $C_{\kappa X}$, $C_{\kappa Z}$, $C_{\omega X}$ and $C_{\omega Z}$. The coefficients $C_{\kappa Y}$ and $C_{\omega Y}$ satisfy the same differential equations as the coefficients $C_{\kappa X}$ and $C_{\omega X}$. }
\label{fig:c_coef}
\end{figure}

The functions $T(\tau)$ and $\mu(\tau)$ shown in Fig.~\ref{fig:Tmu} define the behavior of a hydrodynamic medium whose evolution is decoupled from the spin evolution. The spin degrees of freedom enter here only as trivial degeneracy factors present in the equation of state.

A novel feature of our approach is the possibility to study the evolution of the spin polarization tensor in a given hydrodynamic background. In Fig. \ref{fig:c_coef} we show the time dependence of the coefficients $C_{\kappa X}$, $C_{\kappa Z}$, $C_{\omega X}$ and $C_{\omega Z}$ which define the spin polarization (we have omitted $C_{\kappa Y}$ and $C_{\omega Y}$ as they fulfill the same equations as $C_{\kappa X}$ and $C_{\omega X}$). To compare the relative importance of the coefficients, their initial values have been assumed to be the same. Figure \ref{fig:c_coef} shows a rather weak time dependence of $C_{\kappa X}$, $C_{\kappa Z}$, $C_{\omega X}$ and $C_{\omega Z}$. The strongest time dependence has the coefficient $C_{\kappa Z}$ --- it increases by about 0.1 within 1 fm. We may conclude that the condition that the inclusion of linear terms in $\omega_{\mu\nu}$ is sufficient holds if the initial values of the coefficients $C$ are small and the evolution time is shorter than 10 fm.  
%
\section{Spin polarization of particles at freeze-out}
Let us now demonstrate how our hydrodynamic model can be used to obtain the information about the spin polarization of particles at freeze-out. To achieve this goal we have to define first the freeze-out hypersurface and, subsequently, calculate the average Pauli-Luba\'nski vector of particles with momentum $p$ emitted from this surface. As in the case of the boost-invariant fire-cylinder discussed above, we assume that the freeze-out takes place at a constant value of the longitudinal proper time~$\tau$ and assume the same formula for an element of the freeze-out hypersurface $\Delta\Sigma_\mu$.

By boosting the Pauli-Luba\'nski vector to the rest frame of the particles, we can determine their spin polarization that can be directly compared with the experimental data. In particlular, we can obtain the longitudinal polarization (along the $z$-direction) of particles in their rest frame, which can be studied as a function of transverse-momentum components $p_x$ and $p_y$.
\bigskip
\subsection{Pauli-Luba\'nski vector}
 The phase-space density of the Pauli-Luba{\'n}ski four-vector $\Pi_{\mu}$ is given by the formula \cite{Florkowski:2017dyn}
\begin{equation}
E_p\frac{d\Delta \Pi _{\mu }(x,p)}{d^3 p}
=-\frac{1}{2}\epsilon _{\mu \nu \alpha \beta }\Delta 
\Sigma _{\lambda }E_p\frac{dS^{\lambda ,\nu \alpha }_{\rm GLW}(x,p)}
{d^3 p}\frac{p^{\beta }}{m},
\label{PL1}
\end{equation}
 where $p^\lambda$ is the particle four-momentum. We introduce the parametrization of the particle four-momentum $p^\lambda$ in terms of the transverse mass $m_T$ and rapidity $y_p$,
\beq
p^\lambda &=& \left( m_T\ch(y_p),p_x,p_y,m_T\sh(y_p) \right). \lab{pl}
\eeq
This gives
\begin{equation}
p^{\lambda }U_{\lambda}= m_T\cosh\left(y_p-\eta \right)\lab{PU}   
\end{equation}
and
\begin{equation}
\Delta \Sigma _{\lambda }p^{\lambda }
= m_T\cosh\left(y_p-\eta \right)dx dy\, \tau d\eta. \lab{SIGP}   
\end{equation}

The phase-space density of the GLW spin tensor can be rewritten as~\cite{Florkowski:2018ahw}
\beq
E_p \f{dS^{\lambda , \nu \a }_{\rm GLW}}{d^3p} &=&\frac{{\cosh}(\xi)}{(2\pi)^3 m^2} \, e^{-\beta \cdot p} p^{\lambda } \left(m^2\omega ^{\nu\a}+2 p^{\delta }p^{[\nu }\omega ^{\a ]}{}_{\delta } 
\right).  \label{eq:SGLW22} \nn \\
\eeq
Consequently, using \rf{eq:SGLW22} in \rf{PL1} we can define the total (integrated over the freeze-out hypersurface) value of the PL vector for particles with momentum $p$,
\begin{equation}
E_p\frac{d\Pi _{\mu }(p)}{d^3 p} = -\f{ \cosh(\xi)}{(2 \pi )^3 m}
\int
\Delta \Sigma _{\lambda } p^{\lambda } \,
e^{-\beta \cdot p} \,
\tilde{\omega }_{\mu \beta }p^{\beta }. \lab{PDPLV}
\end{equation}
The contraction of the dual polarization tensor and four-momentum, appearing at the end of the right-hand side of \rf{PDPLV},  gives a covariant four-vector with the components

\begin{widetext}
\begin{equation}
\tilde{\omega }_{\mu \beta }p^{\beta }=\left[
\begin{array}{c}
\phantom{-}\left(C_{\kappa X} p_y-C_{\kappa Y} p_x\right)\sinh (\eta)+\left(C_{\omega X} p_x+C_{\omega Y} p_y\right)\cosh (\eta)+C_{\omega Z} m_T\sinh \left(y_p\right)\\ \\
 \phantom{-}C_{\kappa Z} p_y-C_{\omega X} m_T \cosh \left(y_p-\eta \right)-C_{\kappa Y} m_T \sinh \left(y_p-\eta \right) \\ \\
 -C_{\kappa Z} p_x-C_{\omega Y} m_T \cosh \left(y_p-\eta \right)+C_{\kappa X} m_T\sinh \left(y_p-\eta \right) \\ \\
-\left(C_{\kappa X} p_y-C_{\kappa Y} p_x\right)\cosh(\eta )-\left(C_{\omega X} p_x+C_{\omega Y} p_y\right)\sinh(\eta)-C_{\omega Z} m_T\cosh\left(y_p\right) \\
\end{array}
\right]\,.\lab{OP}
\end{equation}
\end{widetext}

The structure of the last two equations indicates that the total PL vector can be expressed by a combination of the modified Bessel functions. Indeed, straightforward but rather lengthy calculations lead to the expression
\begin{widetext}
\begin{equation}
E_p\frac{d\Pi _{\mu }(p)}{d^3 p}=C_1 K_{1}\left( \hat{m}_T \right)\left[\begin{array}{c}
-\CHI\Big[ \left(C_{\kappa X} p_y-C_{\kappa Y} p_x\right)\sinh (y_p)+\left(C_{\omega X} p_x+C_{\omega Y} p_y\right)\cosh (y_p)\Big] -2 C_{\omega Z} m_T\sinh \left(y_p\right)  \\ \\
 - \big(2 C_{\kappa Z} p_y  -\CHI C_{\omega X} m_T\big)\\ \\
\phantom{-} 2 C_{\kappa Z} p_x  +\CHI C_{\omega Y} m_T \\ \\
\phantom{-} \CHI\Big[\left(C_{\kappa X} p_y-C_{\kappa Y} p_x\right)\cosh (y_p)+\left(C_{\omega X} p_x+C_{\omega Y} p_y\right)\sinh (y_p)\Big] +2 C_{\omega Z} m_T\cosh \left(y_p\right) \\
\end{array}
\right],
\label{totPL}
\end{equation}
\end{widetext}
where $\CHI\left( \hat{m}_T \right)=\left( K_{0}\left( \hat{m}_T \right)+K_{2}\left( \hat{m}_T \right)\right)/K_{1}\left( \hat{m}_T \right)$, $\hat{m}_T=m_T/T$,  and the coefficient $C_1$ is given by the formula
\begin{equation}
C_1=\frac{\pi R^2  \cosh(\xi) \tau m_T}{(2 \pi )^3 m}
\end{equation}
with $R$ being the radius of our system at the freeze-out.
%
%
\subsection{Polarization per particle}
In the next step we have to calculate the average PL vector, {\it i.e.}, the ratio of the total PL vector defined by~\rfn{totPL} and the momentum density of all particles (i.e., particles and antiparticles). The latter is defined by the formula
\beq
E_p\frac{d{\cal{N}}(p)}{d^3 p}&=&
\f{4 \cosh(\xi)}{(2 \pi )^3}
\int
\Delta \Sigma _{\lambda } p^{\lambda } 
\,
e^{-\beta \cdot p} \,.
\eeq
The integration over space-time rapidity and transverse space coordinates yields
\beq
E_p\frac{d{\cal{N}}}{d^3 p}&=&8 m C_1 K_{1}\left( \hat{m}_T \right).
\eeq

The average spin polarization per particle $\langle\pi_{\mu}(p)\rangle$ is obtained by the  expression~\cite{Florkowski:2018ahw}
\beq
\langle\pi_{\mu}\rangle=\frac{E_p\frac{d\Pi _{\mu }(p)}{d^3 p}}{E_p\frac{d{\cal{N}}(p)}{d^3 p}}.
\eeq
One can notice that the coefficient $C_1$ cancels out in this ratio, hence $\langle\pi_{\mu}\rangle$ does not depend explicitly on the chemical potential of the system (which is a consequence of the classical statistics used in this work).
%
\subsection{Boost to the particle rest frame (PRF)}
 In the local rest frame of the particle, polarization vector $\langle\pi^{\star}_{\mu}\rangle$ can be obtained by using the canonical boost \cite{Leader:2001}. Using the parametrizations $E_p=m_T\ch(y_p)$ and $p_z=m_T\sh(y_p)$ and applying the appropriate Lorentz transformation one finds
\begin{widetext}
\beq
\langle\pi^{\star}_{\mu}\rangle&=&-\frac{1}{8m }\left[\begin{array}{c}
0 \\ \\
\left(\frac{\sh(y_p) p_x}{m_T \ch(y_p)+m}\right)\left[\CHI\left(C_{\kappa X} p_y-C_{\kappa Y} p_x\right)+2 C_{\omega Z} m_T  \right] +
  \frac{ \CHI \,p_x \ch(y_p)  \left(C_{\omega X} p_x+C_{\omega Y} p_y\right)}{m_T \ch(y_p)+m}\!+\!2 C_{\kappa Z} p_y  \!-\!\CHI C_{\omega X}{m}_T \\ \\
\left(\frac{\sh(y_p) p_y}{m_T \ch(y_p)+m}\right)\left[\CHI\left(C_{\kappa X} p_y-C_{\kappa Y} p_x\right)+2 C_{\omega Z} m_T  \right] + \frac{\CHI \,p_y \ch(y_p)  \left(C_{\omega X} p_x+C_{\omega Y} p_y\right)}{m_T \ch(y_p)+m}\!-\!2 C_{\kappa Z} p_x \!-\!\CHI C_{\omega Y}{m}_T \\ \\
 -\left(\frac{m\ch(y_p)+m_T}{m_T \ch(y_p)+m}\right)\left[\CHI\left(C_{\kappa X} p_y-C_{\kappa Y} p_x\right)+2 C_{\omega Z} m_T  \right]
-\frac{\CHI \,m\,\sh(y_p) \left(C_{\omega X} p_x+C_{\omega Y} p_y\right)}{m_T \ch(y_p)+m} \\
\end{array}
\right] \nn\\ 
\lab{PLVPPLPRF}
\eeq
\end{widetext}
As expected, the time component of the four-vector  $\langle\pi^{\star}_{\mu}\rangle$ vanishes, since we should have $\langle\pi^{\star}_{\mu}\rangle p^\mu_\star = \langle\pi^{\star}_{0}\rangle m =0$ (in the particle rest frame). We also note that $\langle\pi_{\mu}\rangle \langle\pi^{\mu}\rangle$ is a Lorentz invariant quantity. It can be shown that $\langle\pi^{\mu }\rangle\langle\pi_{\mu}\rangle=\langle\pi _{\star}^{\mu }\rangle\langle\pi^{\star}_{\mu}\rangle$.
%
%

%
\begin{figure*}[t]
      \centering
		\subfigure[]{}\includegraphics[width=0.32\textwidth]{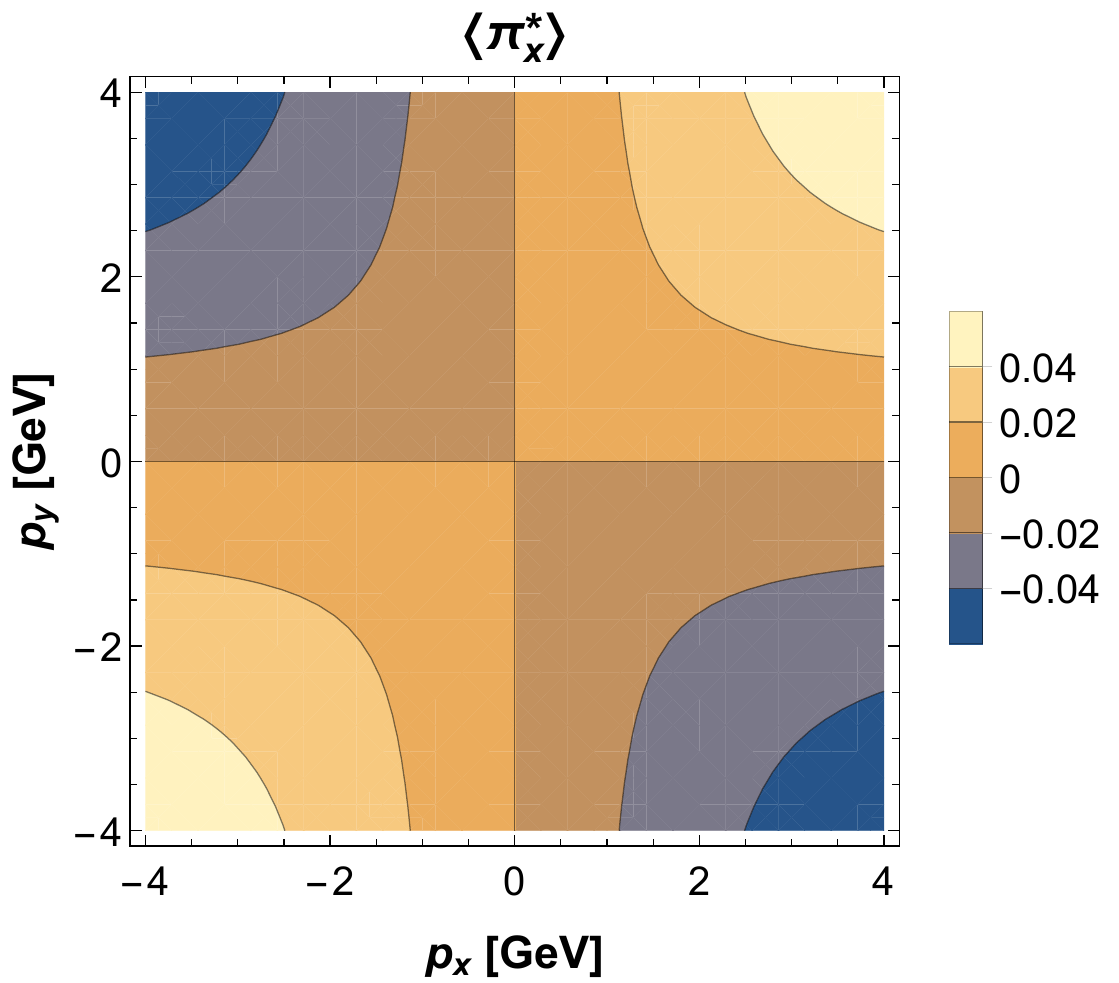}
		\subfigure[]{}\includegraphics[width=0.32\textwidth]{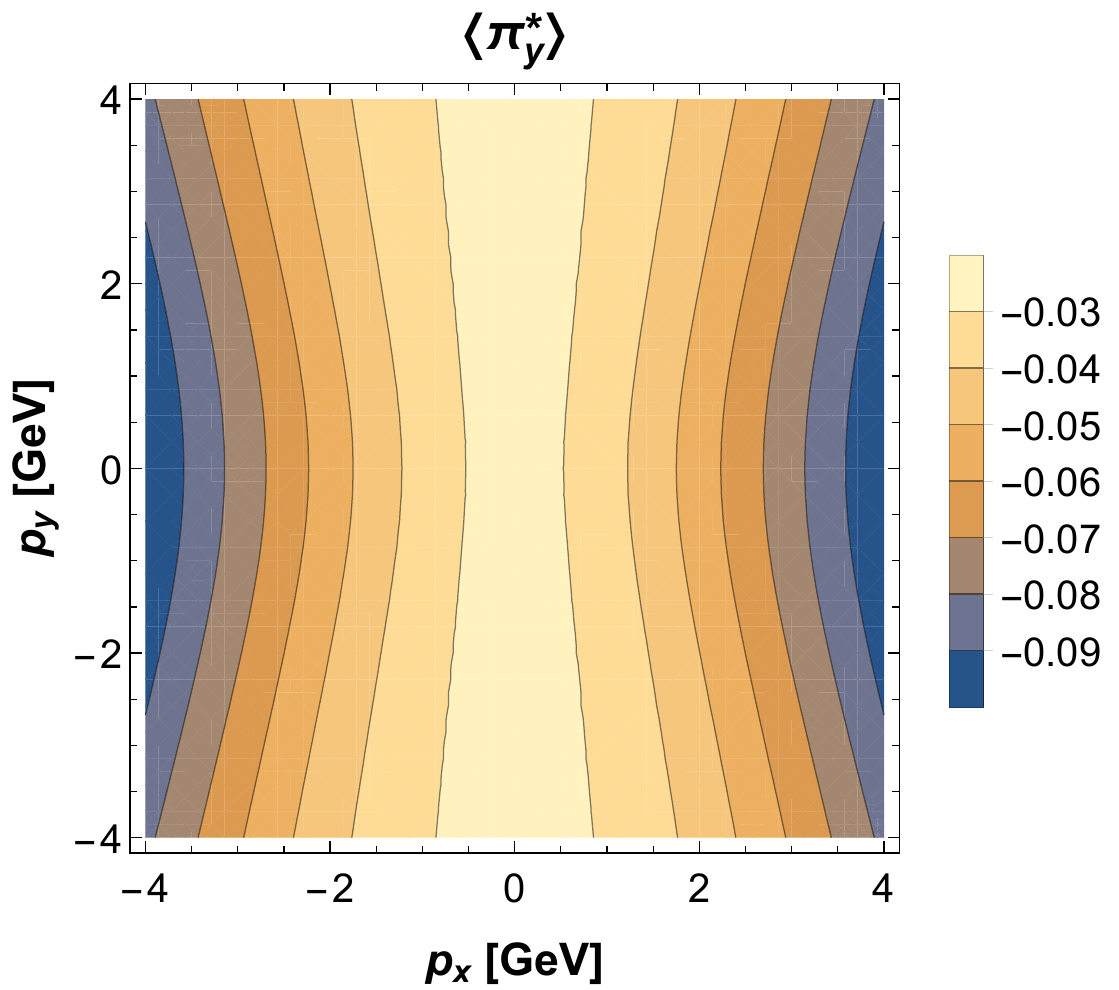}
		\subfigure[]{}\includegraphics[width=0.32\textwidth]{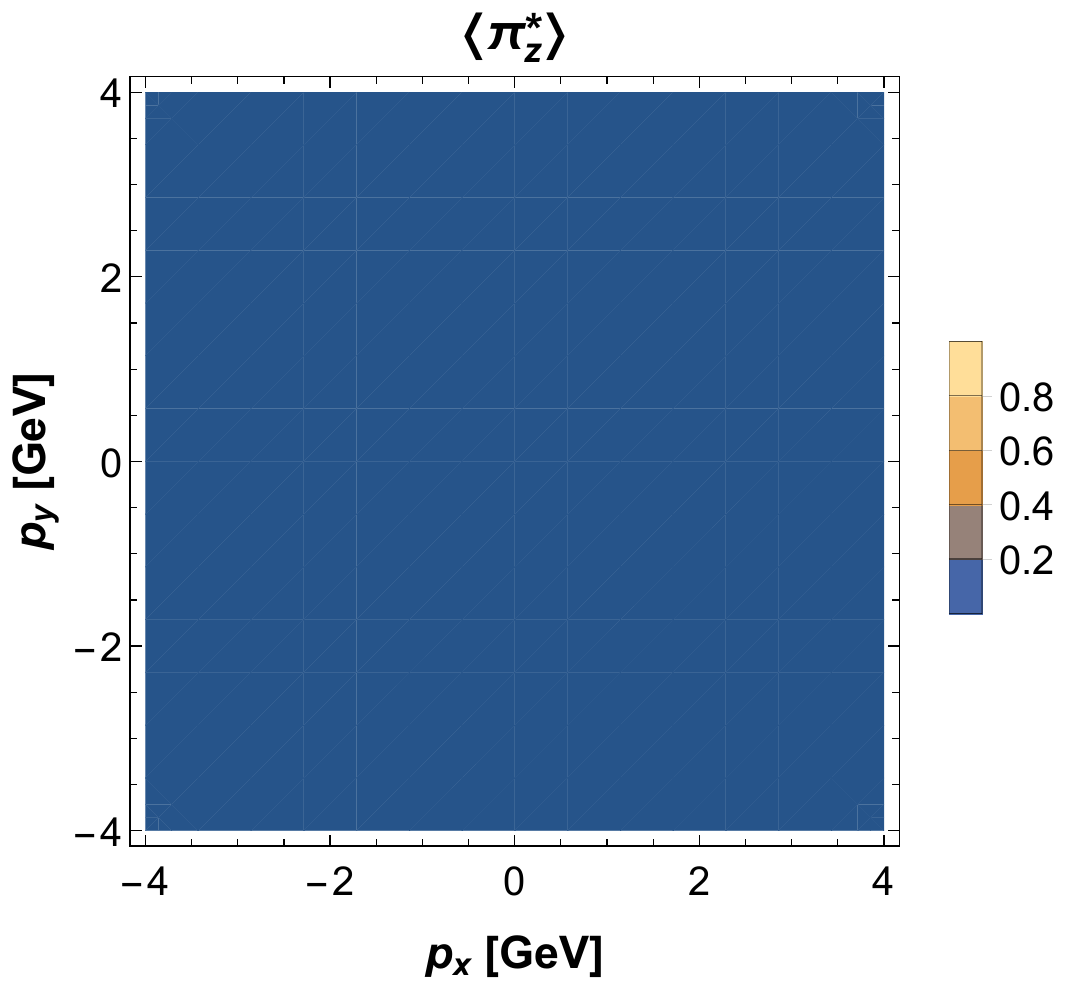}
		\caption{Components of the PRF mean polarization three-vector of $\Lambda$'s. The results obtained with the initial conditions $\mu_0=800$~MeV,  $T_0=155$~MeV, $\Cv_{\kappa, 0} = (0,0,0)$, and $\Cv_{\omega, 0} = (0,0.1,0)$ for $y_p=0$.} 
	 \label{fig:polarization1}
\end{figure*}
\subsection{Approximate expressions}

Since most of the measurements of the spin polarization are done at midrapidity, it is useful to consider particles with $y_p = 0$. Moreover, since the mass of the $\Lambda$ hyperon is much larger than the considered by us values of temperature, $\hat{m}_T \gg 1$, we may use the approximation  $\chi(\hat{m}_T) \approx 2$. Consequently, in this case we obtain a compact expression
\beq
\langle\pi^{\star}_{\mu}\rangle&=&-\frac{1}{4m}\left[\begin{array}{c}
0 \\ \\
\frac{ p_x \left(C_{\omega X} p_x+C_{\omega Y} p_y\right)}{m_T+m}+ C_{\kappa Z} p_y - C_{\omega X}{m}_T \\ \\
\frac{p_y \left(C_{\omega X} p_x+C_{\omega Y} p_y\right)}{m_T+m}- C_{\kappa Z} p_x -  C_{\omega Y}{m}_T \\ \\
- \left(C_{\kappa X} p_y-C_{\kappa Y} p_x\right) -  C_{\omega Z} m_T 
 \\
\end{array}
\right]. \nn\\
\lab{PLVPPLPRFapp1}
\eeq

Introducing the three-vector notation for the polarization vector $\langle \piv^* \rangle = (\langle\pi^{\star1}\rangle, \langle\pi^{\star2}\rangle, \langle\pi^{\star3}\rangle) \equiv (\langle\pi^{\star}_x\rangle, \langle\pi^{\star}_y\rangle, \langle\pi^{\star}_z\rangle)$, and for the coefficient functions $C$, namely 
\beq
\Cv_\kappa &=& (C_{\kappa X}, C_{\kappa Y}, C_{\kappa Z}), 
\label{Ckappa} \\
\Cv_\omega &=& (C_{\omega X}, C_{\omega Y}, C_{\omega Z}),
\label{Comega}
\eeq
we can rewrite \rf{PLVPPLPRFapp1} as 
\beq
\langle \piv^* \rangle = -\frac{1}{4m} \left[
E_p \Cv_\omega - \pv \times \Cv_\kappa - \frac{\pv \cdot \Cv_\omega}{E_p + m} \pv
\right],
\eeq
where we should use $\pv = (p_x, p_y, 0)$. We thus see that for particles with small transverse momenta the polarization is directly determined by the coefficients $\Cv_\omega$. Moreover, since the coefficient functions $\Cv_\omega$ and $\Cv_\kappa$ depend on the freeze-out time in different way, see Fig.~\ref{fig:c_coef}, both the length and direction of the mean polarization three-vector $\langle \piv^* \rangle$ depend on the evolution time. This result may be interpreted also as a change of the polarization during the system expansion.
\section{Momentum dependence of polarization}
Equation \rfn{PLVPPLPRF} allows us to calculate different components of the polarization three-vector as functions of the particle three-momentum. In order to perform such calculations we have to use the values of the thermodynamic parameters and the coefficients $C$ at freeze-out. They can be obtained from the hydrodynamic calculations described in the previous sections. 

One usually argues, that the total angular momentum during the original collision process has only an orbital part which is perpendicular to the reaction plane and negative (the direction of the angular momentum three-vector is opposite to the direction of the $y$-axis). After the collision, some part of the the initial orbital angular momentum can be transferred to the spin part~\cite{Voloshin:2004ha}.  Having this physics picture in mind, we assume that the initial spin angular momentum considered in our calculations has the same direction as the original angular momentum. This is achieved by assuming that $C_{\omega Y}(\tau=\tau_0) > 0$ and other $C$ coefficients are all equal to zero.

The numerical results for this case are shown in Fig.~\ref{fig:polarization1} where we used the initial conditions $\mu_0=800$~MeV,  $T_0=155$~MeV, $\Cv_{\kappa, 0} = (0,0,0)$, and $\Cv_{\omega, 0} = (0,0.1,0)$. Having in mind the measurements done at midrapidity, the results are obtained for $y_p=0$ ($p_z=0$). The three panels of Fig.~\ref{fig:polarization1} show the three components of the polarization three-vector ($\langle\pi^{\star}_x\rangle$,  $ \langle\pi^{\star}_y\rangle$, and $ \langle\pi^{\star}_z\rangle$) as functions of the transverse momentum components $p_x$ and $p_y$.  

As expected, the $ \langle\pi^{\star}_y\rangle$ component (see the middle panel) is negative, which reflects the initial spin content of the system. Because of the assumption $y_p=0$, the longitudinal component is zero (see the right panel). Finally, the component $\langle\pi^{\star}_x\rangle$ shows a quadrupole structure (see the left panel, where the sign changes sequentially throughout the quadrants).

Our results presented in Fig.~\ref{fig:polarization1}  (and other results obtained with different initial conditions, not shown in this work) cannot reproduce an experimentally observed quadrupole structure of the longitudinal polarization. This is a consequence of the symmetries assumed in our simple hydrodynamic model. We note that the hydrodynamic models that use a direct connection between spin polarization and thermal vorticity lead to a quadrupole structure, however, with the opposite sign of the effect as compared to the experimental data~\cite{Karpenko:2016jyx}. The quadrupole structure of $ \langle\pi^{\star}_z\rangle$ appears in connection with the elliptic deformation of the system in the transverse plane and formation of the elliptic flow \cite{Voloshin:2017kqp}. Since our approach assumes homogeneity in the transverse plane, we are not able to reproduce this feature. Our model calculations shown in Fig.~\ref{fig:polarization1} yield a quadrupole structure of the $\langle\pi^{\star}_x\rangle$ component, however, with the sign different from that obtained in the hydro calculations~\cite{Karpenko:2016jyx}.

Clearly, the incorporation of the spin dynamics in fully (3+1)D hydrodynamic models constructed along the lines presented in this work is necessary to address the problems of spin polarization. The presently observed discrepancies between the data and hydrodynamic calculations using the concept of thermal vorticity alone may indicate that there is a place for effects studied in this work.
%
%
\section{Relaxation toward thermal vorticity}
In the hydrodynamic framework defined in this work, it is straightforward to incorporate the effects of dissipative phenomena that can bring the spin polarization tensor $\omega_{\mu\nu}$ to the thermal vorticity $\varpi_{\mu\nu}$. Using the same form of expansion for $\varpi_{\mu\nu}$ as we used for $\omega_{\mu\nu}$, see~\rf{eq:omegamunu}, we can introduce the coefficients $C^{\rm eq}$. The approach of $C$'s toward $C^{\rm eq}$'s can be described by the relaxation-type equations. For example, in the case of the component $C_{\omega Y}$ we can use the equation
\beq 
\frac{dC_{\omega Y}}{d\tau}  &=& \frac{{\cal R}_1}{{\cal P}} C_{\omega Y} + \frac{C_{\omega Y}^{\rm eq}-C_{\omega Y}}{\tau_{\rm eq}}.
\label{rel1}
\eeq
Here, the relaxation time $\tau_{\rm eq}$ is a free parameter (it can be also a function of the proper time).

For the boost-invariant, one-dimensional expansion, the thermal vorticity vanishes, hence, all the coefficients $C^{\rm eq}$ are equal to zero. In this case \rf{rel1} is reduced to the form
\beq 
\frac{dC_{\omega Y} }{d\tau} 
&=& \frac{{\cal R}_1}{{\cal P}} C_{\omega Y} - \frac{C_{\omega Y}}{\tau_{\rm eq}}.
\label{rel2}
\eeq

\begin{figure}[ht!]
\centering
\includegraphics[width=0.45\textwidth]{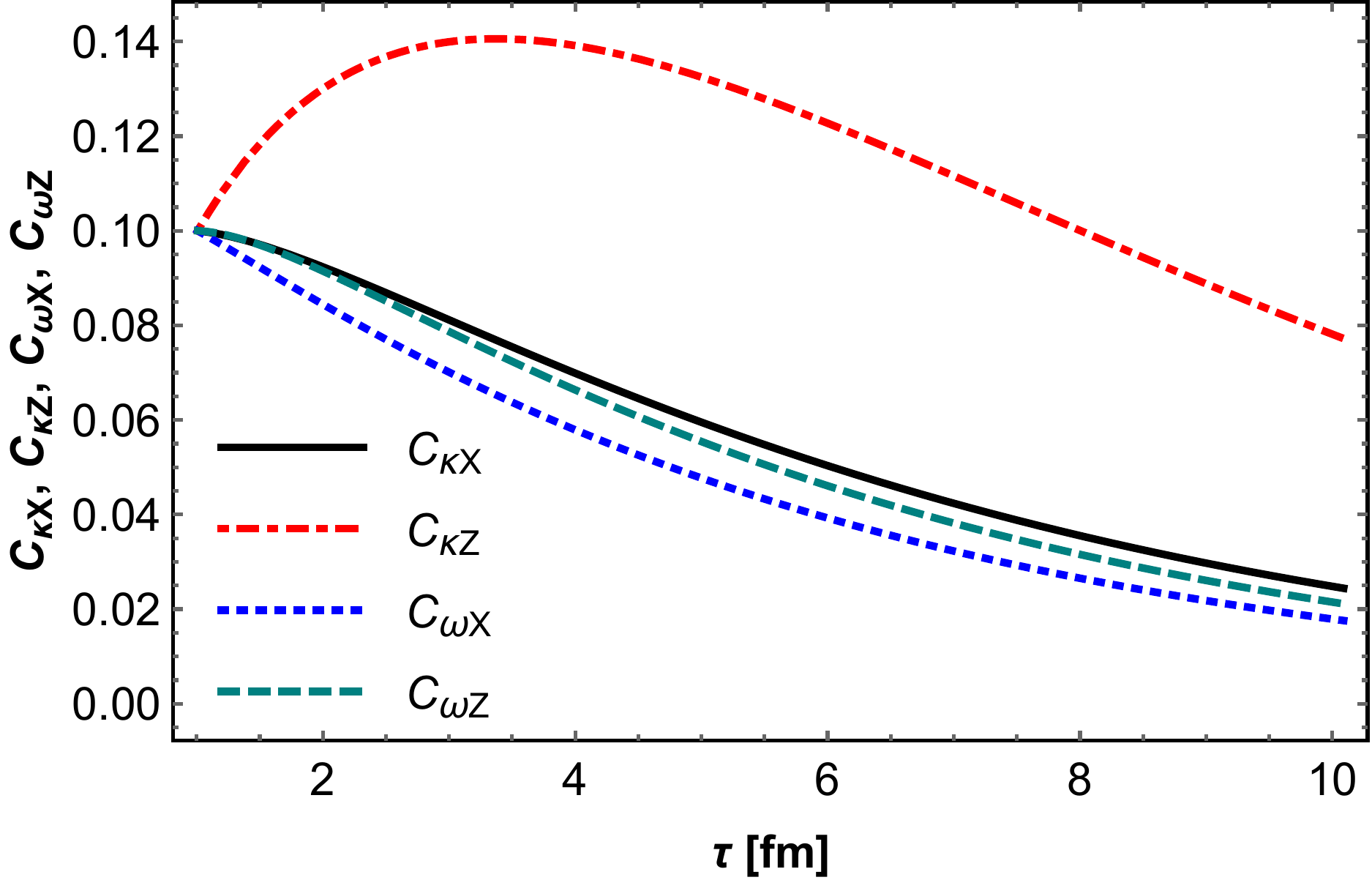}
\caption{Proper-time dependence of the coefficients $C_{\kappa X}$, $C_{\kappa Z}$, $C_{\omega X}$, and $C_{\omega Z}$ in the case where $\omega_{\mu\nu}$ is forced to approach the thermal vorticity $\varpi_{\mu\nu}=0$ with the relaxation time $\tau_{\rm eq}~=~5$~fm.}
\label{fig:c_coef1}
\end{figure}

The numerical results showing the  solution of \rf{rel2} with $\tau_{\rm eq}$~=~5~fm, and the solutions of similar equations obeyed by the coefficients $C_{\kappa X}$, $C_{\kappa Z}$, and $C_{\omega Z}$ are shown in Fig.~\ref{fig:c_coef1}. We see that for the evolution times exceeding $\tau_{\rm eq}$ all the coefficients function approach zero.
\medskip
%
\section{Summary and conclusions}
In this work we have presented first numerical results describing the space-time evolution of the spin polarization tensor in a hydrodynamic boost-invariant background. Our formalism was based on the expressions for the energy-momentum and spin tensors introduced by de Groot, van Leeuwen, and van Weert~\cite{DeGroot:1980dk}, and we considered linear terms in the spin polarization tensor. This procedure allowed us to solve first the standard perfect-fluid equations and subsequently to consider the spin evolution in a well defined hydrodynamic background. 

Our results demonstrate that six scalar functions, which uniquely define a boost-invariant spin polarization tensor, evolve independently. Their proper-time dependence is rather weak, which allows for a consistent treatment of the linear terms. The results of the hydrodynamic calculations can be further used to determine the spin polarization of the particles on the freeze-out hypersurface. 

We have studied in more detail the case where the initial spin angular momentum of the system is described by a vector perpendicular to the reaction plane (identified herein with the $y$-axis). We have demonstrated that the initial direction of polarization is directly reflected in the spin polarization of the particles formed at freeze-out.
\begin{acknowledgments}
The authors thank Sergei Voloshin for clarifying discussions during the Hirshegg 2019 workshop. This research was supported in part by the Polish National Science Center Grant No. 2016/23/B/ST2/00717. 
\end{acknowledgments}
\bibliography{pv_ref}{}
\bibliographystyle{utphys}

\end{document}